\documentclass{cpcauth}

\usepackage{amssymb}
\usepackage{amsmath}
\usepackage[dvips]{graphicx}

\newcommand{\ten}[1]{\boldsymbol{#1}}
\newcommand{\gvec}[1]{\boldsymbol{#1}}
\newcommand{\at}[1]{\arrowvert_{#1}}
\renewcommand{\phi}{\varphi}

\newcommand{\Na}{\bar{N}}

\newcommand{\Tr}{\operatorname{Tr}}
\newcommand{\figref}[1]{Fig.~\ref{#1}}

\begin{document}

\begin{frontmatter}

\title{Stretching of polymers in a turbulent environment}

\author[label1]{Bruno Eckhardt}
\author[label1]{, Jochen Kronj\"ager}
\author[label2,label1]{, and J\"org Schumacher}

\address[label1]{Fachbereich Physik, Philipps-Universit\"at Marburg,
35032 Marburg, Germany}
\address[label2]{Department of Mechanical Engineering, Yale University, 
New Haven, CT, 06520-8284, USA}

\begin{abstract}
The interaction of polymers with small-scale velocity gradients
can trigger a coil-stretch transition in the polymers. We analyze this
transition within a direct numerical simulation of shear turbulence with
an Oldroyd-B model for the polymer. In the coiled state the 
lengths of polymers are distributed algebraically with an
exponent $\alpha=2\gamma-1/De$, where $\gamma$ is a characteristic
stretching rate of the flow and $De$ the Deborah number. In the 
stretched state we demonstrate that the length distribution of the
polymers is limited by the feedback to the flow. 
\end{abstract}

\begin{keyword}
Polymers \sep turbulence \sep coil-stretch transition 
\PACS 
47.50.+d 
\sep
46.35.+z 
\sep
47.27.Eq 
\end{keyword}
\end{frontmatter}

\section{Introduction}
\label{}

Dilute solutions of long polymers show a fascinating variety of
unusal flow phenomena \cite{bird1,joseph}. Recent effects include
the formation of vortex pairs in viscoelastic Taylor-Couette flow
\cite{Groisman1,Graham,Lange} and a phenomenon
called elastic turbulence \cite{Groisman4}. 
While many phenomena in the laminar regime can
be understood within simple constitutive equations \cite{bird1,joseph},
the interplay with turbulence has remained less clear.
In particular, the drag reduction by minute amounts of 
long polymers in turbulent flows \cite{tom49} awaits explanation.
Various experiments
have established that drag reduction comes about when the relaxation
time of the polymers is comparable to the fluctuation times in the
velocity field \cite{lumley,berman}. 
And it is known that a long polymer undergoes a
transition to an uncoiled state when exposed to sufficient strain
\cite{chu,zimmermann}.
Our aim here is to study the interaction of a polymer with 
a turbulently fluctuating flow, in particular the transition 
from the coiled to the stretched state as the internal 
relaxation rate becomes comparable to the velocity gradients,
and to compare to the theoretical predictions of Balkovsky {\it et al.}
\cite{bal00}. Specifically, they predict an algebraic distribution
for the trace of the configuration tensor with an exponent that
depends linearly on the Deborah number and an internal stretching rate
of the flow. Moreover, if the strain is so strong as to uncoil
the flow, this process comes to a stop through its effect on
small-scale fluctuations of the flow.

\section{Model and numerical implementation}


We model the polymer by an upper convected Maxwell fluid with 
a single relaxation time $\lambda$ \cite{max67}. The polymer 
is characterized
by the conformation tensor $c_{ij}$. 
Within a dumb-bell model \cite {bird1} this tensor can be
formed from the end-to-end distance vector $\vec{R}$ as
$c_{ij} = \langle R_i R_j\rangle$ where $\langle\cdots\rangle$ 
denotes a thermal average.
In the coiled state the polymers 
are spherical and $c$ can be normalized so that 
$c_{ij}^{\rm eq}=\delta_{ij}$ and $\Tr c=3$. In the Maxwell model
the relaxation of the polymers is assumed to be linear, so that 
the dynamical equations become
\begin{eqnarray}
\label{eq.v.olb-conf}
  \frac{D}{Dt}c_{ij} 
  - c_{ik}\partial_k v_j -  c_{jk}\partial_k v_i
  = - \frac{c_{ij}-c_{ij}^{\text{eq}}}{\lambda}\, ,
\end{eqnarray}
with $\lambda$ as the time constant and $D/Dt$ as the convective derivative.
The polymers are assumed to be diluted in a Newtonian solvent
so that for the flow properties also the Newtonian shear viscosity
has to be included. This then defines an Oldroyd-B model \cite{old}.
The  Oldroyd-B model is able to describe two main features of viscoelasticity,
namely normal stress differences and stress relaxation, and has been
succesfully applied in the study of vortex pairs in Couette-Taylor flow
\cite{Lange}.
The main
caveat of this model, the divergence of the conformation
tensor when the shear rate of the flow exceeds
$1/\lambda$ is usually overcome with the FENE model \cite{bird1,lsm}. 
However, as predicted by Balkovsky {et~al.} \cite{bal00} and demonstrated below the 
extension of the polymer is stabilized through the feedback 
on the flow once it begins to expand. 

We want to study the turbulence of a flow bounded by two
surfaces and driven by a volume force that maintains a constant mean
shear gradient $S$ (see Fig.~1). This gives an external length
scale $L$ (distance between surfaces) and a velocity scale
$U$ (twice the velocity difference across the gap).\footnote{The unusual 
appearance of a factor 2 is connected with the dimensionless shear rate 
$S=0.5$ used in the code.}
 The Reynolds number then
is $Re=UL/\nu$ with $\nu$ being the kinematic viscosity, and
the Deborah number (characterizing the polymer relaxation) 
is $De=\lambda U/L$. The dimensionless Navier-Stokes and 
Oldroyd-B equations become
\begin{align}
  \label{eq.m.NSE}
  \partial_t v_i + (v_k\partial_k) v_i & =   - \partial_i p 
  + \frac{1}{Re} \partial_{jj} v_i + \partial_j \tau^p_{ij} + f_i\,,\\
  \label{eq.m.OLB}
  \partial_t c_{ij} + (v_k \partial_k) c_{ij} & =  
  c_{ik}\partial_k v_j + c_{jk}\partial_k v_i
  + \frac{c_{ij}-\delta_{ij}}{De}\,.
\end{align}
 
The relation of stress and conformation tensor is given by
\begin{equation}
  \label{eq.m.tc}
  \tau^p_{ij} = \frac{s}{De\,Re}(c_{ij} - \delta_{ij})\,.
\end{equation}
The parameter $s$ in (\ref{eq.m.tc}) models the strength of the
feedback of the polymers on the turbulent flow.
Together with the condition of incompressibility,
\begin{equation}
  \label{eq.m.divv}
  \partial_i v_i = 0\,,
\end{equation}
and a prescribed external force $f_i$,  the set of equations 
\eqref{eq.m.NSE} --  \eqref{eq.m.tc} form a
closed system for the variables $v_i$, $c_{ij}$ and $p$.

The advantage of writing the Oldroyd-B equation in terms of the conformation
tensor $\ten{c}$ instead of the polymer stress tensor $\ten{\tau}^p$ is that
even for $s=0$, when  the polymer has no influence on the velocity field and
$\tau^p_{ij} = 0$, it is still possible to monitor $\ten{c}$ and thus the
stretching and rotation of the polymer by the flow.

\begin{figure}
  \begin{center}
    \includegraphics{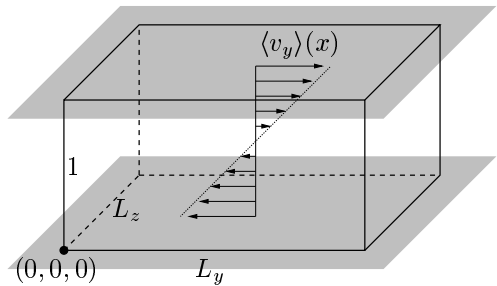}
  \end{center}
  \caption{\label{fig.m.geom}Model geometry: The flow is restricted to the
volume between the shaded planes, with no-flux, free-slip boundary conditions at
the  planes. It is periodically continued in the $y$- and $z$-direction.}
\end{figure}

The model geometry (see \figref{fig.m.geom}) resembles planar Couette flow,
i.e.\ flow between infinite, parallel plates, but with free-slip (or
stress-free) boundary conditions along the planes instead of no-slip boundary
conditions. 
With these boundary conditions all fields can be expanded in Fourier series
and efficient algorithms can be used. In our dimensionless units the box 
has dimensions $\Omega = [0,1]\times[0,L_y]\times[0,L_z]$.

While any quantity is taken to be periodic in the $y$- and $z$-direction,
boundary conditions in the $x$-direction are somewhat more 
complicated and must
be formulated for each quantity separately. For the velocity, they are
\begin{gather}
  \label{eq.m.bcv}
  \begin{aligned}
    v_x\at{x=0,1} = 0\,, &\quad\quad\text{(no flux)} \\
    \begin{aligned}
      \partial_x v_y\at{x=0,1} &= 0\,,\\
      \partial_x v_z\at{x=0,1} &= 0\,.
    \end{aligned} &\quad\quad  \text{(no stress)}
  \end{aligned}
\end{gather}
This translates for the solvent stress tensor $\tau^s_{ij}$
into
\begin{equation}
  \label{eq.m.bctaus}
  \begin{aligned}
    \tau^s_{ij}\at{x=0,1} &= 0\quad
    \text{for } ij = xy, xz, yx, zx,  \\
    \partial_x \tau^s_{ij}\at{x=0,1} &= 0 \quad
    \text{for } ij = xx, yy, zz, yz, zy.
  \end{aligned}
\end{equation}
The boundary conditions on $\tau^p_{ij}$ and 
$c_{ij}$ must be the same as on $\tau^s_{ij}$,
\begin{equation}
  \label{eq.m.bcc}
  \begin{aligned}
    c_{ij}\at{x=0,1} &= 0\quad
    \text{for } ij = xy, xz, yx, zx,  \\
    \partial_x c_{ij}\at{x=0,1} &= 0 \quad
    \text{for } ij = xx, yy, zz, yz, zy.
  \end{aligned}
\end{equation}

Instead of considering the quantities $v_i$ and $c_{ij}$ over $\Omega$ and 
with boundary conditions \eqref{eq.m.bcv} and \eqref{eq.m.bcc}, one can 
extend them to $2\Omega := [-1,1]\times[0,L_y]\times[0,L_y]$ by 
reflection at the $x=0$-plane. The equations of motion 
\eqref{eq.m.NSE} and \eqref{eq.m.OLB} are invariant under this 
transformation. Along the boundary of $2\Omega$, any quantity 
then obeys periodic boundary conditions and can be expanded in 
a three-dimensional Fourier series. Mirror symmetry and boundary 
conditions restrict modes in the $x$-direction to either sines or cosines.

The simulation code is actually written for mixed Fourier ($y,z$) and 
sine/cosine ($x$) expansion coefficients. The  reality condition, 
relations between coefficients resulting from incompressibility 
and de-aliasing by the 2/3 rule \cite{canuto} lead to a further reduction in the 
number of independent variables. In the end one is left with
\begin{equation}
\label{eq.s.diffdim}
(4+8\Na_x) + (5+8\Na_x)(\Na_z+\Na_y+\Na_y\Na_z)
\end{equation}
independent coefficients, where $\Na_i=2N_i/3$ 
for a spatial grid of $(N_x+1)\times  N_y\times N_z$ points.


The numerical algorithm employed for time integration is a 5(4)-Dormand-Prince 
scheme \cite{numrec}
with fixed step size. This scheme uses five evaluations
of the right hand side to
calculate 
an approximation for $\gvec{x}(t+\Delta t)$,
\begin{align}
  \gvec{x}(t+\Delta t) & = \sum_{i=0}^5 \alpha_{6i}K_i + \mathcal{O}(\Delta
t^6)\,,\\
\text{where}\;\;  K_j &= F(\gvec{x}_j, t_j)\,,\notag\\
  \gvec{x}_j &= \gvec{x}(t) + \Delta t \sum_{i=0}^{j-1}\alpha_{ji}K_i\,,\notag\\
  t_j &= t + \gamma_j \Delta t\notag \,,
\end{align}
with coefficients $\alpha_{ij}$ and $\gamma_{i}$ chosen to minimize the error
accumulated over $\Delta t$ \cite{numrec}.
Then, using the 
derivative $F(\gvec{x}(t+\Delta t), t+\Delta t) \approx F(\gvec{x}_6,
t_6)$ (which can be kept for the next integration step), a second 
approximation 
\begin{equation}
  \gvec{x}'(t+\Delta t) =  \sum_{i=0}^6 \alpha'_{6i}K_i + \mathcal{O}(\Delta
t^5)
\end{equation}
of lower order can be obtained. The difference  
$\delta_i =x_i - x'_i$ is used as an estimate for the integration error.
In standard applications this error is then used to adapt the step
size. However, in a turbulent flow with its ever present fluctuations,
variations in this step size are small. So instead we monitor this
error estimate by calculating a combination of relative and absolute error,
\begin{equation}
  \text{Err} = \max_i \frac{|\delta_i|}{|x_i| + \epsilon}\,. 
\end{equation}
With $\epsilon=10^{-6}$ an accuracy of $\text{Err}<10^{-5}$ was maintained 
throughout the integration.

The volume force $f_i$ was chosen so as to maintain the mean shear profile
(dimensionless form)
\begin{align}
  \label{eq.s.lamprofile}
  v^0_y(x) & = \frac{4S}{\pi^2}\,\sum_{m=1}^{\text{M}}
\frac{\cos\left((2m-1)\pi x\right)}{(2m-1)^2}
\end{align}
as an approximation to a linear profile $v^0_y(x)=S(x-1/2)$. 
In the simulations, $M=5$, and the Fourier components contained
in this mean profile were kept fixed. Thus the force is dynamically 
adjusted to maintain the prescribed mean shear profile.


One way to initialize the program is to use a random velocity
perturbation on the laminar profile and the conformation tensor
for the laminar profile. In order to obtain a turbulent state the 
perturbation has to be about as large as the laminar profile.
In turns out, however, that even for moderate Deborah numbers 
$De\approx 0.1,\ldots, 1$ this leads to an extreme overshooting
of the conformation tensor and a breakdown of the integration routine.

As a remedy to this we used initial conditions from a turbulent
Newtonian flow with equilibrium configuration tensor, 
$c_{ij}^{\text{eq}} = \delta_{ij}$. The system was then allowed
to relax to a statistically stationary turbulent state for a small
value of the Deborah number. One of these states was then taken
as initial condition for a simulation at a higher $De$. 
Proceeding in this manner at each change of $De$ the 
system relaxed to a statistically stationary state within
a few time steps and no further numerical integration 
problems were encountered.

\begin{figure}
  \begin{centering}
    \includegraphics[scale=0.4]{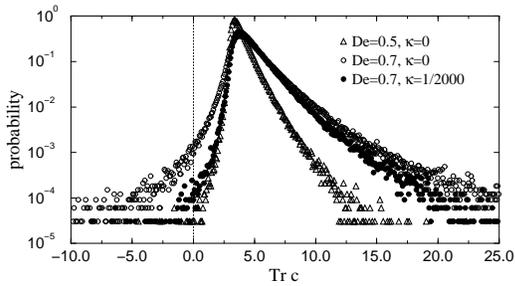}\\
  \end{centering}
  \caption{\label{fig.r.positrace}Probability density functions (PDFs) 
    of $\Tr \ten{c}$ in a turbulent flow at different $De$. There are
    hardly any regions with negative $\Tr c$ for $De=0.5$ (open
    triangles) and a significant fraction for $De=0.7$ (open circles).
    Adding an artifical diffusivity of $\kappa=1/2000$ removes the
    negative values and leaves the positive distribution essentially
    unchanged (filled circles).  Parameters: $Re=1000$, $S=\partial_x
    \langle v_y \rangle = 0.5$ and $s=0$. Geometry: $L_y=4\pi$,
    $L_z=2\pi$, $N_x=32$, $N_y=128$, $N_z=64$. The PDFs are averaged
    over 10 snapshots taken at time intervals $\Delta t=5$.}
\end{figure}

Another numerical problem for Oldroyd-B simulations is connected
with the fact that
the conformation tensor $c_{ij}$ is by definition positive definite,
and that this property is preserved by the Oldroyd-B equations
\cite{joseph} but not in the numerical representation.
Through the various Fourier transforms and
truncations, this positivity can be lost.  If that happens the
Oldroyd-B model can develop Hadamard instabilities, and lose its
evolutionary character \cite{dup86}. Thus, if regions with negative
$c$ appear, numerical instabilities can arise and the integration can
blow up \cite{ber99}. \figref{fig.r.positrace} shows that such regions
indeed do accur, but remain limited to a small fraction of the whole
domain, provided $De$ is not too large.
%

As a partial cure to this problem,
Beris \textit{et~al.}{}\cite{ber99,ber95,ber96} introduce an artificial
stress diffusion $-\kappa\Delta\ten{c}$ in the equation of motion for
the conformation tensor, where $\kappa$ can be chosen so that it
stabilizes the numerics but has negligible influence on the results.
Although for the small values of $De$ in \figref{fig.r.positrace} no
numerical instability seems to occur, the introduction of a small
$\kappa$ does reduce the probability for negative $\Tr \ten{c}$, while
the impact on positive values is much less significant.
In the following, calculations for $De\leq 0.5$ are without artificial
stress diffusivity ($\kappa =0$). The number of points with $\Tr
\ten{c} < 0$ is then negligible. For $De > 0.5$, a small
$\kappa=1/2000$ is employed to reduce negative tails of PDFs, and for
$D=2.0$ it is actually necessary to set $\kappa=1/1000$ to avoid
numerical instability. In both cases, a
certain probability for negative $\Tr\ten{c}$ is still unavoidable.

The simulation code is written in Fortran 90 (except the special FFT
routines for Cray T90 hardware). It is based on the code
used previously \cite{sch00} for the same model without polymer. Most
simulations were done on a Cray T90 vector machine at the John von
Neumann Institute for Computing at the Research Centre
J\"{u}lich. A typical run with a resolution of $N_x=32$, $N_y=128$,
$N_z=64$ over $50$ time units costs roughly $26500$ CPU seconds and
needs 62 megawords of memory.

\section{\label{sec.r.lebedev}Statistics of polymer elongation}



Balkovsky {\it et al.} \cite{bal00} discuss the behaviour of the PDF for
the polymer extension $R = (\Tr\ten{c})^{1/2}$. For a passive polymer,
they predict a power law decay of the tails ($R\rightarrow \infty$),
as long as the (dimensionless) relaxation time of the polymer $De$ is
smaller than the inverse shear rate $1/(2\gamma)$ of the underlying
turbulent flow,
\begin{equation}
\label{eq.r.lebedev1}
  \mathcal{P}(R) \sim R^{-\alpha-1}\,.
\end{equation}
The exponent $\alpha$ depends on $De$ and $\gamma$. For relaxation times near
the inverse Lyapunov exponent, $\alpha$ is given in linear approximation by
\begin{equation}
\label{eq.r.lebedev2}
  \alpha \sim \frac{1}{De} - 2\gamma\,.
\end{equation}
For Deborah numbers smaller than $De_{cr}=1/(2\gamma)$ 
the exponent $\alpha$ is positive and
\eqref{eq.r.lebedev1} actually describes an algebraic decay. This corresponds to
the majority of the polymer molecules being near their equilibrium elongation. 
However, for values above the critical Deborah number $De_{cr}=1/(2\gamma)$
the exponent $\alpha$ becomes negative and the probability density function
for $R$ is no longer normalizable.  At this point the
assumption of a passive polymer becomes unphysical. Balkovsky {\it et al.}
\cite{bal00} 
then argue
that the feedback from the polymer to the flow field limits the extension, such
that $\mathcal{P}(R)$ decays again for large $R$. The majority of the molecules
is then stretched to some finite value of $R$, at which $\mathcal{P}(R)$ is 
maximal. The contribution of the polymer stress $\tau^p$ to the right hand side
of the Navier-Stokes equation should then be of the same order of magnitude as
the advective term $\vec{v}\cdot\nabla\vec{v}$.



The quantity directly accessible from the simulations is $\Tr\ten{c} = R^2$. The
PDFs for $R^2$ and $R$ are related through 
\begin{equation}
  \mathcal{P}(R)\,dR = \mathcal{P}(R^2)\,dR^2 = 2R\,\mathcal{P}(R^2)\,dR\,.
\end{equation}
A power law distribution for one quantity implies one
for the other, and the exponents are related by
\begin{equation}
\mathcal{P}(\Tr\ten{c})\sim \left(\Tr\ten{c}\right)^{-\beta}\Leftrightarrow
\mathcal{P}(R) \sim R^{-\alpha-1}\,,
\end{equation}
with $ \alpha = 2(\beta - 1)$.

\begin{figure}
  \begin{center}
    \includegraphics[scale=0.43]{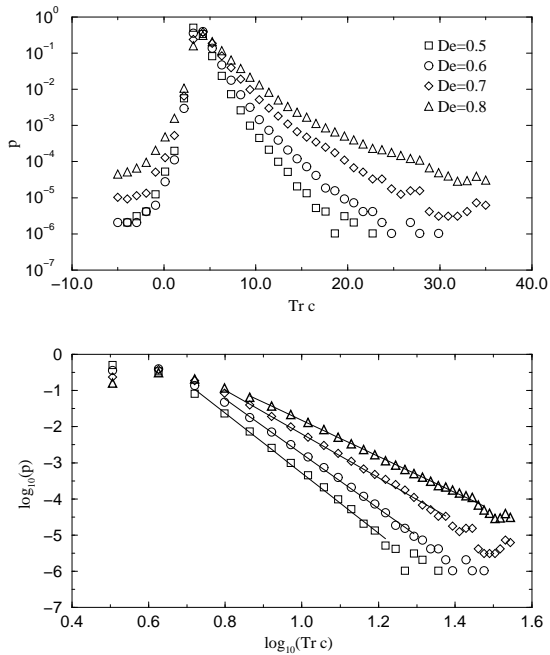}
  \end{center}
\caption{\label{fig.r.lebedev1}PDFs of $\Tr \ten{c}$ in turbulent flow for
different values of $De$. The straight lines in the lower frame follow
from algebraic fits over the ranges over which they are shown.
Every data point corresponds to one bin.
Parameters: $Re=1000$, $\partial_x \langle v_y \rangle
= 0.5$ and $s=0$. $\kappa=1/2000$ for $De\geq 0.6$. Geometry: $L_y=4\pi$,
$L_z=2\pi$, $N_x=32$, $N_y=128$, $N_z=64$. The initial condition was the same
for all $De$. The PDFs are averaged over 10 snapshots taken at time intervals
$\Delta t=5$.}
\end{figure}

\figref{fig.r.lebedev1} demonstrates these power tails in PDFs obtained from
turbulent simulations for different $De$. The exponents $\beta$ are obtained
from the slope in a log-log plot (lower graph). The region where the power law
actually holds starts when $\mathcal{P(\Tr\ten{c}})$ drops below about $10^{-1}$
and extends right up to the statistical accuracy limit for larger $\Tr\ten{c}$. 

\begin{figure} 
    \begin{center}
      \includegraphics[scale=0.40]{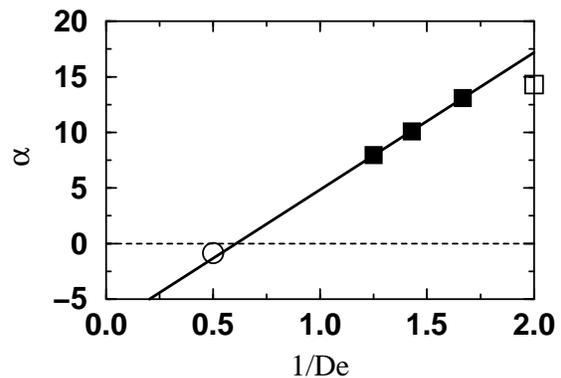}
    \end{center}
\caption{\label{fig.r.lebedev2}Exponents extracted from \figref{fig.r.lebedev1}
($De \leq 0.8$, squares) and the equivalent for 
($De=2.0$, $s=0.01$, open circle). 
The regression line derived from the solid squares intersects with 
$\alpha=0$ for $1/De=0.61 = 2\gamma$. Note that the open circle
for the situation of uncoiled polymers also lies on the line.}
\end{figure}

 From the exponents determined in \figref{fig.r.lebedev1}, the Lyapunov
exponent $\gamma$ (and with it, the critical Deborah number) can be
obtained by plotting $\alpha$ as a function of $1/De$, as in
\figref{fig.r.lebedev2},
and extrapolating the linear relationship to
$\alpha=0$. Note that the three points corresponding to $De=0.6$, $0.7$ and
$0.8$ fit the line surprisingly well.  The deviation for $De=0.5$ could be
attributed to the fact that only this simulation is done with $\kappa=0$. Also,
$De=0.5$ is farthest from the (extrapolated) critical Deborah number. 


\begin{figure}
  \begin{center}
    \includegraphics[scale=0.45]{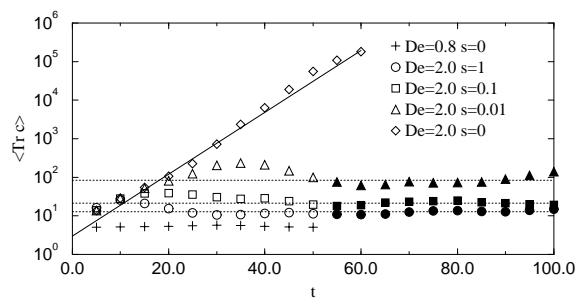}
  \end{center}
\caption{\label{fig.r.lebedev4} Time evolution of $\langle \Tr\ten{c}\rangle$
for $De > De_{cr}$. Without feedback through the flow the polymer
length diverges. With feedback ($s\ne0)$ it stabilizes.
The horizontal dotted lines mark the time average of the solid
points; these averages decrease with increasing coupling strength $s$.}
\end{figure}

 From the data on the exponents one can extract that for $De$ larger than
 $1/0.61$
the polymer distribution without feedback is no longer normalizable,
the polymers go into a stretched state. This stretching is terminated
by their feedback on the flow, as we will now demonstrate. To this
end we analyze the time evolution of $\langle\Tr\ten{c}\rangle$ 
in \figref{fig.r.lebedev4}.
The trace starts at the equilibrium value $\Tr\ten{c}=3$ at $t=0$.
For $De<De_{cr}$ it increases and then stabilizes at a finite value,
indicating a certain stretching of the polymer. For $De>De_{cr}$ it
increases monotonically, if there is no feedback to the flow ($s=0$),
but stabilizes at finite values for $s\ne0$. The larger the feedback,
the smaller the mean values.

In an additional calculation for $De=2.0$ and $s=0$ 
(see \figref{fig.r.lebedev4}),   $\langle \Tr\ten{c}\rangle$ grows more 
or less exponentially, as expected, with
a rate $d\ln\langle \Tr\ten{c}\rangle/dt \approx 0.18$. The difference
between $1/De_{cr}=0.61$ and $1/De=0.5$ gives 
a growth rate of about $0.1$. The difference is perhaps connected to the use of different averages in the calculation of the exponents.

\section{Conclusions}
The full numerical simulations for an Oldroyd-B constitutive equation 
in a turbulent shear flow support the calculations of Balkovsky {\it et~al.}
and demonstrate and algebraic distribution of polymer lengths below
the coil-stretch transition. Above the transition the calculations
also show that the potentially infinite stretching of the polymers is
limited by their modulation of the small-scale flow properties.

\section*{Acknowledgement}
This work was performed as part of the European Community
Research Training Network HPRN-CT-2000-00162.
The numerical simulations were done on a Cray T-90 at the John von
Neumann Institute for Computing at the
Research Centre J\"ulich and we are grateful for their support.

\end{document}